# Coupling of individual quantum emitters to channel plasmons

Esteban Bermúdez-Ureña[1], Carlos Gonzalez-Ballestero[2], Michael Geiselmann[1]†, Renaud Marty[1], Ilya P. Radko[3], Tobias Holmgaard[4], Yury Alaverdyan[5], Esteban Moreno[2], Francisco J. García-Vidal[2,6], Sergey I. Bozhevolnyi[3] and Romain Quidant[1,7]

[1]ICFO–Institut de Ciencies Fotoniques, Mediterranean Technology Park, 08860 Castelldefels (Barcelona), Spain

[2]Departamento de Física Teórica de la Materia Condensada and Condensed Matter Physics Center (IFIMAC), Universidad Autónoma de Madrid, ES-28049 Madrid, Spain

[3]Department of Technology and Innovation, University of Southern Denmark, Niels Bohr Allé 1, DK-5230 Odense M, Denmark

[4]Department of Physics and Nanotechnology, Aalborg University, Skjernvej 4A, DK-9220, Aalborg Øst, Denmark

[5]The Nanoscience Centre, University of Cambridge, 11 JJ Thomson Avenue, CB3 0FF Cambridge, United Kingdom

[6]Donostia International Physics Center (DIPC), E-20018 Donostia/San Sebastian, Spain

[7]ICREA – Institució Catalana de Recerca i Estudis Avançats, Barcelona, Spain

† Current address: Ecole Polytechnique Fédérale de Lausanne (EPFL), CH–1015, Switzerland

Correspondence should be addressed to EBU (esteban.bermudez@icfo.es) and RQ (romain.quidant@icfo.es):

## Abstract

Efficient light-matter interaction lies at the heart of many emerging technologies that seek on-chip integration of solid-state photonic systems. Plasmonic waveguides, which guide the radiation in the form of strongly confined surface plasmon-polariton modes, represent a promising solution to manipulate single photons in coplanar architectures with unprecedented small footprints. Here we demonstrate coupling of the emission from a single quantum emitter to the channel plasmon polaritons supported by a V-groove plasmonic waveguide. Extensive theoretical simulations enable us to determine the position and orientation of the quantum emitter for optimum coupling. Concomitantly with these predictions, we demonstrate experimentally that 42% of a single nitrogen vacancy centre emission efficiently couples into the supported modes of the V-groove. This work paves the way towards practical realization of efficient and long distance transfer of energy for integrated solid-state quantum systems.

# Introduction

Hybrid systems consisting of quantum emitters (QEs) coupled to plasmonic waveguides (PWs) have received much attention as building blocks for future quantum plasmonic circuitry platforms[1–11]. These architectures not only provide a fundamental insight into strong light-matter interaction[12–14] or quantum many body physics[15–18] but are also envisioned to enable a variety of applications, such as on-chip generation and routing of single photons[19,20], single-photon transistors[9] or PW-based quantum interferometers[21] among others. To achieve such functionalities, multiple criteria have to be fulfilled simultaneously. Among these criteria, the PW mode propagation length, the decay rate enhancement (Purcell factor) and the QE-PW mode coupling efficiency (β factor) are of utmost importance: their product, normalized by the operation wavelength, defines a figure of merit (FOM) that quantifies the ability of such systems to achieve efficient long-range energy transfer. Practically, achieving a large FOM is exceedingly challenging, since this requires combining an appropriate PW configuration, which should exhibit moderate losses as well as strong mode confinement, with a deterministic coupling of an individual QE to the PW mode.

Considering possible PW configurations, it should be noted that, even though strong mode confinement and large propagation lengths have been achieved with chemically synthesized metallic nanowires[22–25], these structures are hardly suitable for a controllable circuitry design. Top-down lithography-based fabrication techniques could in principle overcome this issue, but the resulting PWs usually suffer from larger losses[26]. An attractive type of PW is the so-called V-groove (VG) channel waveguide, which is among the most promising candidates for developing a planar plasmonic circuitry platform[2,27–34]. The VG waveguides represent hollow V-shaped channels carved in a metal surface that support the propagation of channel plasmon polaritons (CPPs). The CPPs combine the unique properties of subwavelength confinement of the electromagnetic fields near the VG bottom, reasonably long propagation[30,31] and low losses at sharp bends[28]. Apart from conventional photonic circuit components, VGs were recently employed to demonstrate resonant guided wave networks, opening thereby an exciting perspective for designing novel dispersive and resonant photonic materials[34]. Furthermore, recently developed approaches for local CPP excitation via the integration of nanomirror tapers to the VG terminations[35] or butt-coupling with a silicon-on-insulator waveguide excited through a grating coupler[34] provide excellent routes for the CPP in- and out-coupling to free-space propagating light, opening an easy on-chip access to CPP-based plasmonic circuitry. Overall, these remarkable developments place the VG´s performance in close competition with the chemically synthesized nanowires, but the fabrication methods associated with the VGs enables for a flexible and realistic plasmonic circuitry design[36], thus inclining the balance in favour of the VGs as the PW of choice.

Experimental and theoretical investigations of hybrid QE-PW systems have mainly focused on the study of nanowires[22–25,37] to guide surface plasmons excited by semiconductor quantum dots[4,6,18,33,38–41] or nitrogen-vacancy (NV) centres in nanodiamonds (NDs)[19,42–45]. In particular, the implementation of NV centres in such hybrid systems has been motivated by their remarkable characteristics, among which brightness and room temperature stability make them a very promising solid-state single photon source[46]. Additionally, the presence of a spin triplet ground state is extremely appealing for solid-state quantum communication and computing schemes[47,48]. The electronic spin can be optically initialized and read-out, exhibits long coherence times and microwave pulses can be implemented not only to manipulate the spin but also to improve the coherence time by decoupling from the nuclear spin bath[49]. Furthermore, experimental spin-photon entanglement demonstrations over the recent years have strengthen the potential of NV centres as one of the key candidates for quantum registers in a future quantum network realization[48,50].

In this work we demonstrate for the first time coupling of a single NV centre to CPPs supported by a VG channel. Using theoretical analysis and simulations, we first study the behaviour of a dipolar emitter coupled to a VG channel and identify the key features of such a hybrid system. We find optimal coupling for a dipole aligned transversally to the VG long axis and that the vertical position range allowing for an efficient QE-CPP coupling corresponds to dipole positions far away from the VG bottom. This requirement imposed by the CPP field distribution (determined in turn by the VG profile) made the choice of the NV centres in NDs more appealing, since the presence of a diamond shell facilitates locating the NV centre at the appropriate height. To realize experimentally the hybrid NV centre–VG device we utilize state-of-the-art assembling techniques[51–53] to deterministically position a single NV centre inside a VG PW. The coupling of the NV centre emission into the CPP mode is then evidenced by wide field fluorescence imaging. The observation of QE-CPP coupling together with significantly long propagation length is in accordance with our theoretical predictions: the investigated configuration out-performs previous realizations with respect to the proposed FOM[4,43,45].

## Results

**VG channels as a platform for quantum plasmonic circuitry.** The configuration investigated, both experimentally and theoretically, comprises the assembly of a ND, hosting a single NV centre, inside a gold VG PW (Fig. 1a). Upon excitation with a 532 nm laser, the single NV centre will, in the ideal case of perfect coupling, direct all of its emission into the VG-supported CPP modes. The CPP-guided emission propagating along the VG will then outcouple at the VG nanomirror terminations. The corresponding VG structures have been milled with a focused ion beam in a 1.2-µm thick gold layer[29] (see Methods), aiming to produce narrow and deep VGs terminated with tapered nanomirrors (Fig. 1b and insets) for efficient CPP out-coupling[35]. In order to confirm that the fabricated VGs enable sub-diffraction confinement of the electromagnetic energy together with a long propagation within the emission band (~600-800 nm) of NV centres in diamond, numerical simulations were performed[33,54] for a VG with geometrical parameters corresponding closely to those measured from the SEM images (lower inset of Fig 1b). The simulations indicate considerable confinement of the VG-supported CPP electric field, peaking up close to the VG bottom while being practically constant across the VG (Fig. 1c for λ=650 nm). As expected, the electric field lines (black arrows) inside the VG show that the supported mode is TE-polarized[35]. In the case of an infinitely long VG channel, simulations predict an intrinsic CPP propagation length of 4.56 µm when considering the contribution from three wavelengths within a 100 nm range of the NV spectrum (see Methods for details on the simulations).

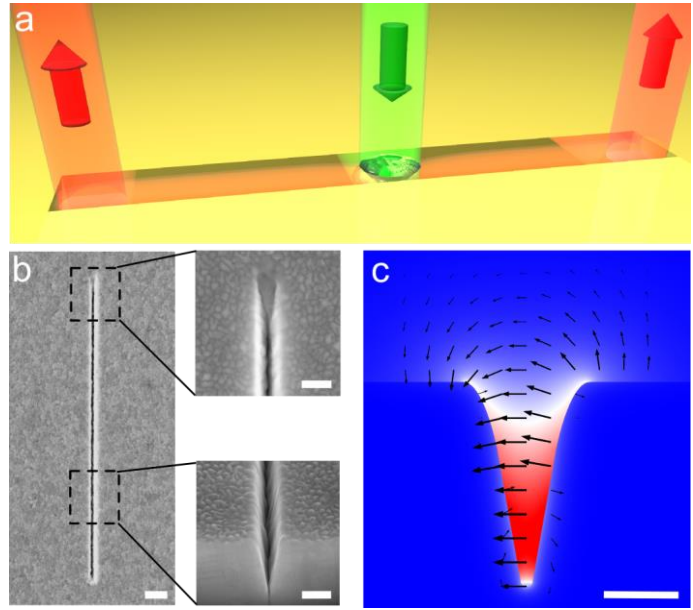

**Figure 1. V-groove channel waveguide platform for quantum plasmonic circuitry.** (a) Schematic of the considered configuration. A single ND hosting a single NV centre is placed inside a VG channel waveguide. Upon excitation with a 532 nm green laser, the NV centre couples, in the ideal case, all of its emission into the VG-supported CPP mode. The channelled emission out-couples from the VG via the tapered nanomirrors at the VG extremities. (b) Scanning electron microscopy (SEM) image of a 10-μm-long V-shaped groove (~315 nm width and ~510 nm depth) fabricated by milling a thick gold film with a focused ion beam. The scale bar is 1μm. The top right inset is a zoomed SEM image from the tapered nanomirror at the VG end. The bottom right inset is a transversal view from a cross-cut of one VG and evidences the V contour of the channel. The scale bars of the insets are 300 nm. (c) Total electric field profile of the VG-supported CPP mode for a wavelength of 650 nm. In the simulations, the VG dimensions (315 nm width and 510 nm depth) have been chosen to reproduce the experimental conditions. The field lines of the electric field are represented by the black arrows. The length of the arrows is proportional to the logarithm of the field intensity at each point. The scale bar is 200 nm.

**Modelling of the QE coupling to a VG supported CPP mode.** The characterization of the hybrid system comprised of the VG presented in Fig. 1 and a single QE (here a dipole emitter) has been carried out in two steps. The first step consisted of a 2D simulation to determine the CPP mode profile supported by an infinitely long VG[33,54](Fig. 1c). In a second step, the 3D problem was tackled to determine the optimum parameters for the position of a single QE inside the VG (see Methods for further details). The Purcell and β-factors, displayed in Fig. 2(a), were calculated using a standard procedure for a bare dipole emitter in absence of the diamond host[33](Fig. 2a). Simulations for three wavelengths within the NV emission spectrum were carried out, namely for 650 nm, 690 nm and 750 nm, and the final values were averaged by weighting the contribution of each wavelength to the NV emission spectrum (Supplementary Figures 1 and 2 show the individual wavelength results). In the most favourable orientation, namely transversal to the VG axis (x-axis in Fig. 2), the β-factor rises up to 68% (black curve in Fig. 2a), so that the maximum QE-CPP mode coupling is achieved at distances between 200 nm and 330 nm from the VG bottom. In this region, the decay rate increases by a factor of 5 as compared to the vacuum decay rate (magenta curve in Fig.2a). When the QE approaches the bottom of the VG, the Purcell enhancement reaches higher values but the coupling is less efficient, as the decay is dominated by ohmic losses[55,56].

Our theoretical results shed light on the importance of controlling the position of a single QE inside such a PW. In nanowire PWs, the optimal distance between a QE and the metallic surface for efficient coupling to the guided modes is in the order of 10 nm[4,5,25,40,42,43]. In contrast, the VG PWs are favourable to QEs located at distances in the order of 50 nm away from the metallic surfaces. The possibility to enhance the Purcell factor far away from the metallic surfaces allows maintaining a low non-radiative decay rate for the QE, which consequently decreases intrinsic losses for the hybrid device and increases the emission decay rates of the single photon source[55,56]. The efficient coupling (large β-factor) of the dipolar source at such positions revealed in our simulations (Fig. 2a) along with long CPP propagation qualifies this platform as a favourable configuration enabling quantum plasmonic circuitry. The NV centres in NDs are ideal candidates to fulfil the conditions required for an efficient coupling with the peculiar modes supported by these PWs as the diamond host acts as a spacer between the NV centre and the metallic surfaces. Noteworthy, this spacing of the shell also alleviates the typical non-radiative relaxation of a QE when being brought close to the metallic surfaces, which usually results in fluorescence quenching[55,56]

To further analyse this hybrid configuration, we considered a 60-nm-radius ND placed inside the VG, lying within the optimum β-factor region depicted by the shaded area, to study the effect of the ND shell on the QE-CPP coupling. The evolution of the field profile along the VG for the QE radiating inside the ND particle (Fig. 2b) demonstrates that indeed, a QE embedded in such a ND shell can efficiently couple to the guided CPP modes, as the field profile far away from the QE (third panel at 1.25 μm) resembles that of the CPP mode. To calculate the β-factor, a different method is required as the translational symmetry of the bare VG is broken after the introduction of the ND sphere. The computation scheme is based on the overlap of the transversal fields emitted by the QE and the CPP mode of the empty VG[54]. The influence of both the source orientation and its vertical position within the ND on the Purcell and β-factors was studied for a fixed position of the ND (Fig. 2c). The presence of the ND shell (Fig. 2c) does not affect considerably the results obtained in the case of a bare dipolar source (Fig. 2a). Indeed, both the Purcell enhancement and coupling efficiency remain drastically reduced for unfavourable dipole orientations (y and z axis). Remarkably, when the QE is oriented along the adequate direction (x-axis in Fig. 2), the coupling between the QE placed at the centre of the ND sphere (dashed line in Fig. 2a) and the VG supported CPP mode is almost as efficient when compared to the bare dipole case, since the β-factor is only reduced to 56% (black curve in Fig.2c) while keeping a moderate Purcell factor of 5.2. The realistic simulations of a QE coupled to the VG-supported CPP mode unambiguously demonstrate that the ND shell preserves the QE-CPP efficient coupling while increasing the emitter´s decay rate (Purcell factor).

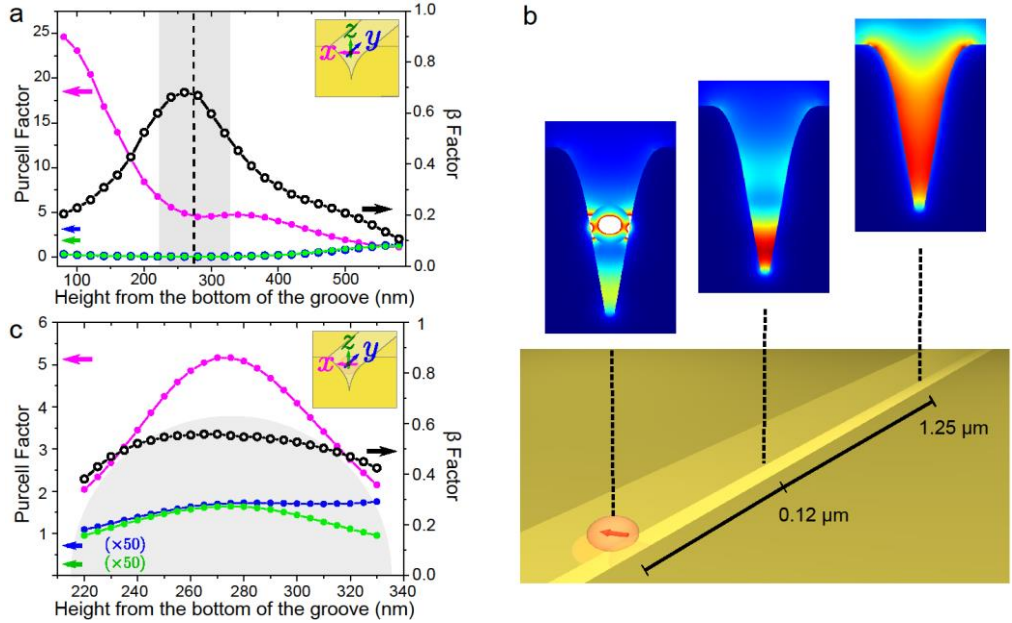

**Figure 2. Simulations of the single QE-CPP mode coupling.** (a) Characterization of the coupling between a bare dipolar emitter and the VG-supported CPP mode. The dipole is placed in the centre of the VG and displaced along the vertical z-axis. The shaded region represents the ND which will be considered inside the VG later on. The Purcell factors for the three orthogonal dipole orientations are displayed with a colour code corresponding to that used in the inset. The black line shows the β-factor for a dipole oriented along the x-axis for which the Purcell factor is enhanced. The β-factor is found to be close to zero for y and z orientations (not shown). b) Normalized electric field maps for three different transversal cuts of the VG. The electric field profile resembles that of the VG-supported CPP mode already at 1.25 μm away from the QE (dipole oriented along the x-axis). The colour scale has been adjusted in each panel for a better view. c) Same quantities as in (a), when the dipolar source is located inside a 60-nm-radius ND sphere lying inside the VG. The β-and Purcell factors reach values of 0.56 and 5.2 respectively.

**Deterministic assembly of a single NV centre-VG hybrid device.** In order to experimentally assemble such a hybrid device, it is important to first identify a bright and stable single NV centre. In our ND solution, typically around 5% of the particles are single NV-NDs. Therefore, the first crucial step is to locate the ND particle featuring the desired single emitter characteristics, so as to prepare the ground for assembling in a truly deterministic fashion an efficient hybrid quantum plasmonic device. To position a single ND inside a VG, two recently developed nano-positioning techniques for QEs were combined, namely the use of electron beam lithography (EBL) based assembly of QEs[51] and the nano-manipulation of individual particles with an atomic force microscope (AFM) tip[52,57].

First, an array of NDs was deposited in a controlled fashion in the vicinity of the VGs. To do so we implemented an EBL based positioning method[51] with the aid of electrostatic self-assembly to attach the NDs to the substrate[53] (see Methods for further details). Confocal fluorescence microscopy scans under a 532 nm green laser excitation allowed us to locate the optically active NDs (Fig. 3c). Furthermore, by implementing a Hanbury-Brown-Twiss (HBT) detection scheme[58], we could identify the fluorescent NDs hosting a single NV centre. It is important to underline that this strategy also guarantees the quantum nature of the plasmonic device since the CPPs can only be launched one by one by this single photon source. We choose one of the measured single NVs ($g^2(t=0)<0.5$ in Fig. 3f) which exhibited both a relatively large fluorescence lifetime (Fig. 3e) as well as a large count rate (bright single photon source), since this combination enables to choose an NV centre presenting a priori a large quantum efficiency [59,60].

In the second step, we used an AFM in tapping mode for visualization, and in contact mode[52,57] to move the ND across the Au film and finally into the VG. Figure 3a shows an AFM image of a VG with the surrounding array of positioned NDs. The red solid circles in Figs. 3a,c indicate the location of the chosen ND containing a single NV centre (see Fig. 3f) and the dashed circles show the position at which we intended to relocate this ND inside the VG. We first transferred the ND to a position close to the VG in order to assess the stability of the NV centre´s emission properties upon the movement. To do so, the confocal scans as well as the lifetime and correlation measurements were repeated for the same ND (green and blue traces in Fig. 3e). The acquisition of similar lifetime values demonstrates the stability of the emission properties of the chosen NV centre under translation of the ND in a homogeneous environment.

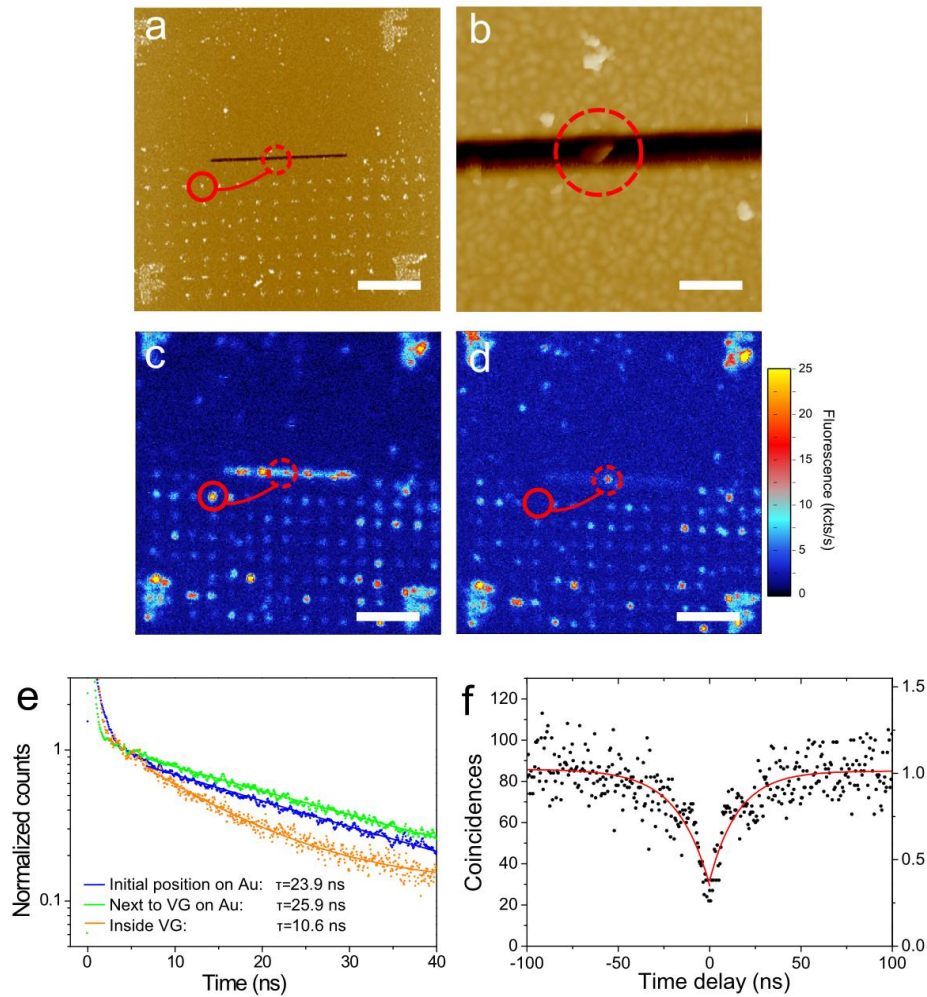

**Figure 3. Deterministic assembly and characterization of a single NV centre inside a VG channel**. (a) AFM image of NDs positioned in the vicinity of a VG. The solid and dashed circles indicate, respectively, the position of one ND with a single NV and the position in the VG where it was positioned with an AFM tip. The scale bar is 5 µm. (b) AFM scan of the ND inside the VG channel. The scale bar is 500 nm. (c)-(d) Confocal fluorescence microscopy images of the VG and surrounding array of NDs before and after AFM manipulation of the ND hosting a single NV centre. The map in (d) evidences the presence of the ND inside the VG. To enhance the contrast of the NV centre fluorescence, this image was acquired when the excitation polarization was set parallel to the VG main axis as this configuration reduces the Au auto-fluorescence of the VG. The scale bars are 5 µm. (e) Lifetime measurements of the selected NV centre under 532 nm green pulsed laser excitation at three different positions: initial and intermediate positions at the Au surface (blue and green) and inside the VG (orange). (f) Second order autocorrelation of the selected ND. The presence of a dip at zero time delay (t=0) shows the quantum nature of the NV. For the selected particle, $g^2(t=0) < 0.5$ unambiguously demonstrates the presence of only one single NV centre inside the ND.

Finally, the ND was moved inside the VG channel as highlighted with the red dashed circle in the AFM image in Fig. 3b. The confocal scan of the same area confirms that the NV centre is indeed located inside the VG (Fig. 3d). It is worth noticing that a clear visualization of the NV centre position within the VG was obtained by setting the excitation polarization parallel to the VG in order to reduce the Au auto-fluorescence from the VG. Indeed, the latter yields considerably stronger signals when excited with a polarization transversal to the VG main axis due to the contribution from gap surface plasmons[61] as well as from wedge surface plasmons[62–64]. The Au auto-fluorescence emission cannot be completely filtered out spectrally as its contribution spreads into the NV centre emission range. Nevertheless this fluorescence process is much faster than that of the NV centre[65], therefore enabling us to distinguish those two processes in time and ultimately determine the NV centre lifetime independently of the Au auto fluorescence process (single exponential fits were performed in the range between 4 ns and 60 ns). We observed a lifetime change from 25.9 ns to 10.6 ns after positioning the ND inside the VG (Fig. 3e), which corresponds to a total decay rate enhancement factor of ~2.44. In order to estimate an experimental Purcell factor one cannot rely on the measurement performed on the same emitter on the Au film, as the latter can support surface plasmon polaritons that can substantially contribute to the measured lifetime. Instead, we compare the value measured inside the VG to the average of the lifetime distribution measured on single NV centres from the same solution deposited on a glass substrate (Supplementary Fig. 3, $\tau=24.2 \pm 7.2$ ns), and obtained an experimental Purcell factor of $2.3 \pm 0.7$. The distribution on glass is consistent with other reports on similar sized nanodiamonds; the longer lifetime as compared to a NV centre in bulk diamond (~11.6 ns) is typically attributed to the reduction in the radiative emission rate caused by the variation in effective refractive index of the surrounding medium for emitters embedded in nanoparticles smaller than the emission wavelength[46,59]. To compare the experimental Purcell factor to theoretical simulations, we calculated the lifetime distribution associated to an isolated ND when considering an ensemble of ND sizes (40 nm-80 nm radii). This distribution was then used to normalize the Purcell factors presented in Fig.2. For the 60-nm-radius ND with its NV centre aligned to the transversal axis of the VG (Fig. 2c), we obtain a modest Purcell factor of $3.0 \pm 0.6$, while for the other orthogonal orientations we get a negligible value. Therefore, this system presents a FOM of 11.1, which is a large value in comparison to previously proposed configurations relying on the use of colloidal nanowires[4,42,45].

**Coupling of a single NV centre to VG supported CPPs.** To unambiguously demonstrate the ability of this quantum plasmonic device to couple the NV centre emission to the CPP mode supported by the VG, we have performed wide-field collection fluorescence imaging around our confocal excitation spot by using an EMCCD camera. Generally, under confocal excitation with a diffraction-limited spot, one observes only the fluorescence emitted from the excitation point. In contrast, with wide field collection, when the excitation spot coincides with the position of the single NV centre inside the VG, additional out-coupling spots located only along the axis of the VG could be observed. The polarization dependence of those spots was evidenced by the wide-field collection fluorescence images for four combinations of excitation and collection polarizations, i.e. combining polarizations parallel and transversal to VG axis (Figs. 4b-e). Additionally, under wide-field illumination, we could appreciate the contour of the VG due to the enhanced auto-fluorescence of the structure with respect to the plain Au film[64] (Fig. 4a).

First, it is interesting to have a close look at the spot brightness coming directly from the ND, referred to as the confocal spot. We find that the confocal spot is brighter for a collection polarization transversal to the VG axis (Fig. 4b-c) compared to the parallel axis collection (Fig. 4d-e), independently of the excitation polarization. This polarized emission is consistent with our previous observation concerning the most likely NV centre orientation, i.e. aligned towards the transversal axis of the VG. The difference in the intensities of the confocal spots for the two excitation polarizations (Figs. 4b-c) is attributed to the enhanced Au auto-fluorescence arising from the gap and wedge plasmons supported by the VG as discussed in the previous section.

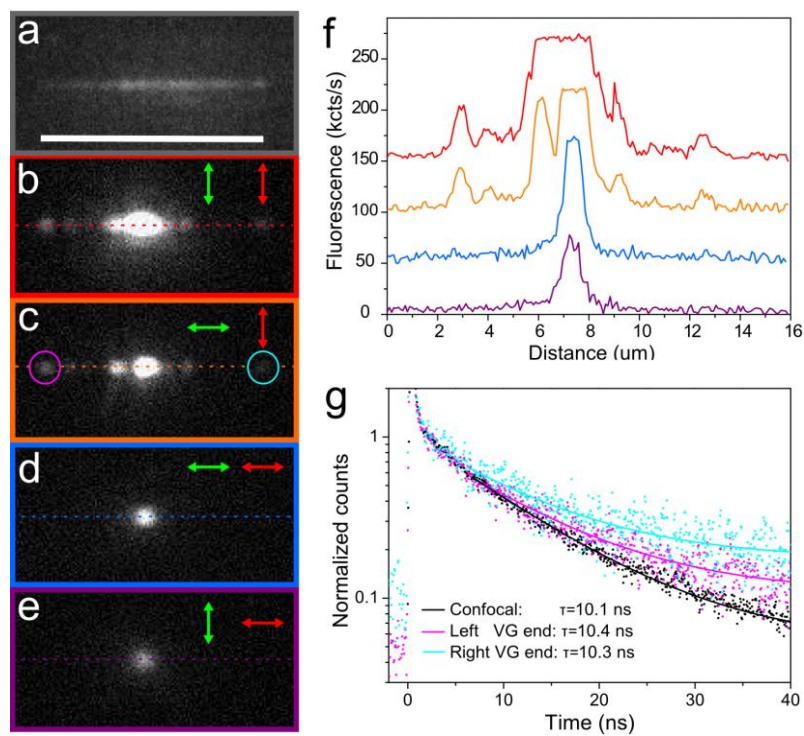

**Figure 4. Coupled emission of a single NV centre to CPPs in a VG.** (a)-(e) Wide field collection fluorescence microscopy images of an assembled NV-VG device. (a) Upon wide-field illumination the contour of the VG can be detected due to the enhancement of the Au auto-fluorescence at the VG location. (b)-(c) Collection polarization (red arrow) set transversal to the VG axis evidences the guided emission out-coupled from the VG tapered mirrors. Additional out-couplers along the VG axis correspond likely to defects in the VG. The excitation polarization (green arrow) was set transversal (b) and parallel (c) to the VG axis, respectively. (d)-(e) Collection polarization set parallel to the VG axis. Only emission coming directly from the excited NV centre is observed for both parallel (d) and transversal (e) excitation polarizations. (f) Fluorescence intensity along the dashed line in (b)-(e) (same colour code as the frames). (g) Lifetime measurements for the polarization configuration of Fig.4(c) at three different collection locations. Direct collection (black) and emission out-coupled from the left (magenta) and right (cyan) VG extremities. Similar lifetime values measured at the three positions confirm that the guided emission corresponds to that of the single NV centre coupling to CPPs.

Considering the simulations presented in Fig.2, a QE having a dipole moment oriented transversally to the VG axis should efficiently couple to the supported CPP mode, given that it sits at the proper height range inside the VG. The out-coupling spots observed experimentally at the extremities of the VG (Figs.4b-c) clearly demonstrate the coupling of the single NV centre emission to the CPP mode and its propagation over a distance of at least ~5 μm along the VG. Additionally, it is clear that the out-coupling spots are only appreciated when the collection is set transversally to the VG main axis. The observation of only a confocal spot in Figs. 4d-e confirms the TE polarized nature of the propagating CPP mode (Fig.1c). To quantitatively analyse the difference between those configurations and highlight the presence or the absence of the out-coupling emission spots, line-cuts of the fluorescence signal (coloured dashed lines in Fig. 4b-e) are plotted in Fig. 4f as a function of the distance along the VG axis. On the line-cuts associated to Figs. 4(b)-(c), the two out-coupling spots at the ends of the VG do not exhibit significant differences. This suggests that the Au auto-fluorescence, which is higher for the confocal spot under transversal excitation, does not couple significantly to the CPP modes responsible for the emission observed at the VG extremities. The main reason why the Au auto-fluorescence is not able to couple to the CPP mode is that its fluorescence is generated at the surface of the VG walls and thus not fulfilling the position requirements for efficient coupling as shown in Fig. 2.

Moreover, some additional out-coupling spots along the VG axis are observed for the transversal collection. Note that their locations are independent of the excitation polarization and correspond likely to defects along the VG that scatter out part of the guided emission. In addition, absence of these spots for the parallel collection indicates that all additional spots in Figs. 4b-c are manifesting the propagation of excited CPPs and validates our conclusions regarding the coupling of the NV centre emission with CPP modes as the only mechanism to transfer energy along the VG. The presence of these defects and the amount of light they scatter out will determine the amount of energy reaching the VG ends, and in turn the effective propagation length of a given device can vary depending on which segment is being considered. The intrinsic propagation length of the CPP mode on the other hand is independent of such scatterers, depends only upon the PW geometry and material composition, and can be well approximated with proper numerical simulations. Experimentally it is challenging to measure *in-situ* the intrinsic propagation length of a given device. Nevertheless we can estimate the performance of nominally equivalent VGs onto which we deposited NDs with multiple NV centres per particle. By analysing the wide field EMCCD images evidencing the coupling to the CPPs we extracted a value of 4.65 ± 0.48 μm, in close accordance to our simulated value for the intrinsic propagation length value (see Supp. Info for details).

To confirm that the observed out-coupling spots at the VG extremities correspond to emission from the NV centre, we performed lifetime measurements under pulsed excitation at three different collection positions while fixing the excitation point at the NV centre (Fig. 4g) for the case of parallel excitation and transversal collection: from the confocal spot (black dots) and from the left and right VG extremities (magenta and cyan dots, respectively). Single exponential fits yield essentially the same lifetime values, confirming that the observed out-coupling spots are indeed resulting from the coupling of the NV centre emission into the CPP mode.

The coupling efficiency (β-factor) of the hybrid system is defined as the ratio between the decay rate into the CPP mode and the total decay rate of the QE. Experimentally the measured lifetime is related to the total decay rate of the QE regardless of the collection position (Fig.4g). Therefore, it is not possible to measure separately the lifetimes associated to the different decay rates of a QE. Instead, we can estimate the experimental β-factor by computing the ratio between the intensities of the out-coupling spots associated with CPP coupled emission (VG ends and other scatterers along the VG) and the sum of all the out-coupling spots, including the confocal spot. To do so, we analysed the intensity counts from non-saturated images taken in absence of a polarizer in the collection channel (Supplementary Figure 5). Correcting the measured intensities to account for the propagation losses, we obtained a β-factor of 0.42 ± 0.03 (Supplementary Table 1). This value is in good agreement with the expected value of 0.56 extracted from the simulations presented earlier (Fig. 2b). The small deviation from the optimum value is reasonable considering that the position and/or orientation of the NV centre are not necessarily ideal in the experiment.

The combination of a β-factor of 0.42 ± 0.03 and a Purcell factor of 2.3 ± 0.7 for our hybrid system, together with a propagation length of 4.65 ± 0.48 µm, leads to a FOM of 6.6 ± 1.5, larger than the value of 4.2 ± 1.9 extracted from the seminal work of Akimov *et al*[4] based on a hybrid colloidal quantum dot-silver nanowire system. Similar values were also obtained for a second device (Supplementary Figure 5), exhibiting a β-factor of 0.41 ± 0.05 and a Purcell factor of 2.1 ± 0.6. The accordance between the experimental results and the simulated values for the ideal NV orientation, suggests that the ND positioned inside the VG hosts a NV centre aligned not far from the transversal axis of the VG. These results places our hybrid system in a good position within the FOM that seeks simultaneously for high coupling efficiency and long propagation lengths while exhibiting a moderate decay rate enhancement, providing a benchmark to compare QE-PW hybrid systems.

**Discussion**

In the field of quantum plasmonics, hybrid systems consisting of a QE coupled to a PW have already been realized. To bridge the gap allowing the integration of such a device onto a functional chip, the proposed configuration tackles two major issues. The first one arises from the dissipative nature of the PWs at visible wavelengths. Here, we chose to face this problem by employing a VG PW supporting CPP modes, which are known to provide relatively low losses with subwavelength confinement as well as flexible and realistic circuit designs. The second issue concerns the choice of QE and its coupling to the CPP mode.

When dealing with nanowire-based PWs, achieving a large coupling efficiency usually requires the QE to be so close to the metal that the non-radiative decay rate starts to contribute significantly. This results in fluorescence quenching, which can make the source dark. Nevertheless, on-chip quantum plasmonic devices require, while dealing with a bright stable source, both a long propagation length and high coupling efficiency that together account for a large FOM. In this context, we demonstrate theoretically and implement experimentally a quantum plasmonic device based on a VG integrated with a single NV centre, hosted inside a ND, which enables to reach a favourable trade-off between these two constraints. The ND shell is used as a natural spacer, providing optimal vertical position for the NV centre and allowing its separation from the metallic surfaces, protecting the QE from non-radiative coupling to the metal and therefore providing a bright and stable single photon source that can efficiently couple to the CPPs. Moreover, theoretical predictions indicate that this ND only weakly perturbs the properties of the CPPs and the QE-CPP coupling. Experimentally, we deterministically assemble NV-VG devices and observe coupling from the single NV centres to the CPP modes of VG channels by combining state of the art positioning techniques. We demonstrate energy transfer from a single QE over a 5 µm distance before out-coupling this energy into free-space propagating light by means of tapered nanomirrors. We obtain a Purcell factor of 2.3 ± 0.7 and a coupling efficiency of 0.42 ± 0.03 together with a propagation length of 4.65 ± 0.48 µm, in close accordance to our theoretical simulations. This confirms a larger FOM compared to previous realizations based on colloidal plasmonic nanowires. Our approach has the additional advantage of a top-down fabrication technique that can enable realistic and functional plasmonic circuitry. Furthermore our measurement and analysis methods are fundamental to understand and take advantage of the polarization dependent performance of such hybrid devices.

In the future, similar hybrid systems can be assembled and characterized thanks to the combined methods presented here. To improve the efficiency of the device, near-infrared single photon sources such as silicon vacancy centres in ND could be used.[66] One can also aim to improve the coupling efficiency by controlling the NV dipole orientation with respect to the CPP mode field lines (transversal to the VG axis), for example by implementing optical trapping techniques.[67] Finally, it should be borne in mind that VGs can specifically be designed (and fabricated) to contain smoothly connected sections with different VG profiles so as to enhance the NV-CPP coupling at a very narrow VG section with extremely tight mode confinement[68] while preserving large CPP propagation lengths at wider VG sections.[69] Among the next steps might be to develop an integrated quantum plasmonic function based on this hybrid system. For instance, an on-chip optical transistor or a Mach-Zender PW interferometer could be realized.

## Methods

**Sample fabrication** The VGs are prepared by milling a 1.2-µm-thick gold layer deposited on a silicon substrate by means of a focused ion beam (FIB). They are ~315 nm wide and ~510 nm deep, have an opening angle of ~24°, a length of 10 µm, and are terminated with ~650 nm long width-constant tapers. During fabrication of such a taper, the dwell time was reduced continuously along the groove (resulting in decreased depth) with the groove width being constant. In this case, the VGs are tapered only in one dimension, i.e., in depth. Since the width is kept constant, the bottom of the waveguide transforms smoothly from V-shape into a flat surface, thereby forming a triangular-shaped nano-mirror.

The monocrystalline diamonds were purchased from Microdiamant (MSY 0-0.2 µm) and were post-processed with cleaning and filtering steps as described in Ref. 67. To position the nanodiamonds we first spin coated a 120-nm thick layer of poly (methyl methacrylate) (PMMA) (Microchem, 950k 4 wt % diluted 3:1 with trichloromethane) and baked at 90 ºC for 15 min on a hot plate. Note that the baking temperature was set lower than the typical 175 ºC to avoid deformation of the VG structure due to Au annealing. We then patterned an array of 200-nm holes by electron beam lithography (EBL). The pattern was positioned in close proximity to the VG by using a set of alignment marks which were milled during the FIB process. The sample was then developed in a 1:3 solution of Methyl isobutyl ketone (MIBK) and isopropanol (IPA) for 45 s. Next, we placed a drop of a solution containing a positively charged polyelectrolyte (poly) diallyldimethylammonium (PDDA) (Sigma Aldrich, $M_W$ 200000–3500000, 2 wt % in Milli-Q water, Millipore) onto the structures to perform the electrostatic assembly. After 5 min of incubation the sample was rinsed with de-ionized water in order to remove excess of PDDA and blow dried in a $N_2$ stream. We then placed a drop of a nanodiamond solution (particle radii in the range of 40-80 nm) and let incubate for 30 min. To avoid an excess of particles during the lift-off we first removed the solution drop with a pipette and rinsed the sample in IPA. Finally lift-off in acetone at 55 ºC during 30 min followed by rinsing in IPA and blow-drying in a $N_2$ stream leaves us with the VG surrounded by the designed pattern of nanodiamonds. The nanodiamonds used for the propagation length estimation were purchased from Adamas Nanotechnologies and had nominally 15 NVs per particle. They were randomly deposited onto a substrate with several VGs in order to obtain several coupled devices within the same sample.

**Device characterization.** Experiments were performed at room temperature on a home-built fluorescence microscopy setup. For the confocal scans, excitation was done with a 532 nm CW laser (Ventus) with excitation power in the order of 85 µW. The laser was filtered with a dichroic mirror and additional filters were used to filter the fluorescence of the NV centres and reduce the Au fluorescence contribution below 633 nm. For the confocal measurements a collection channel with a 25-µm pinhole was implemented. After passing through the pinhole the signal was sent to two APD detectors. Using a Picoharp300 system (Picoquant), autocorrelation measurements where performed to correlate the emitted photons from the identified emitters. Also, lifetime measurements were performed with this Picoquant system while exciting an NV centre with a 532 nm pulsed laser (LDH-FH, Picoquant) and collecting the signal with one of the APDs.

To perform the wide field fluorescence measurements we implemented an electron multiplying charged coupled device (EMCCD) camera (Hamamatsu) accessible through another collection channel by means of a flip mirror. To achieve a wide-field illumination and identify the contour of the VGs [Fig. 4(a)], we focused our 532 nm CW excitation laser onto the back focal plane of the objective by means of a flip lens. In all the experiments excitation and collection polarization where adjusted by combining a linear polarizer and a quarter-waveplate to rotate the excitation polarization of the green laser, while the collection polarization was filtered with a linear polarizer.

**Theory simulations.** The field distributions were calculated numerically with the Finite Element Method, using the COMSOL multi-physics tool. The analysed VG in the numerical simulations has the following main geometric properties: an aperture angle of 24º at the bottom, with a 15 nm radius rounded bottom, a depth of 510 nm, and a width of 315 nm. The surrounding dielectric material is assumed to be air ($\varepsilon=1$), whereas the gold is described by an experimentally fitted Drude-Lorentz formula[70]. The NV centre is described as a current $I_0$ oscillating along a very short rectilinear segment of length $l= 2$ nm, with the frequency $\omega_0 = 2\pi c/\lambda_0$, $c$ being the vacuum light velocity. All the simulations described hereafter were performed for three values of $\lambda_0$: 650 nm, 690 nm and 750 nm, and the final results for the intrinsic propagation length, Purcell and Beta factors where weighted against the NV spectrum contribution at each wavelength. When the diamond host is present, it is modelled as a dielectric sphere of radius $r = 60$ nm and dielectric constant $\varepsilon_d = 5.737$. In these simulations, where this sphere is allowed to fall inside the VG, its centre lies at 234 nm below the aperture. The simulation domain is a prism of dimensions 1.3x1.3x2.2 µm, all the faces being terminated in perfectly matched layers. The VG is assumed to point along the $y$ axis.

As a first step, we carried out a simulation of the bare VG in order to characterize the CPPs. By solving a standard 2D eigenvalue problem we were able to obtain both the modal fields and the complex propagation constant $\kappa$. From the latter, we determined the propagation length at each wavelength. We obtained values of 3.4 μm, 4.7 μm and 6.6 μm for the wavelengths of 650 nm, 690 nm and 750 nm respectively, yielding an average weighted value of 4.56 μm.

In order to obtain the Purcell Factor we used the following general expression for a radiating point dipole: $P_F = \frac{6\pi c}{\omega_0} \mathbf{u}_\mu Im[\overleftrightarrow{G}(\mathbf{r}_\mu, \mathbf{r}_\mu, \omega)]\mathbf{u}_\mu$, where $\mathbf{r}_\mu$ and $\mathbf{u}_\mu$ are the position of the dipole and the unit vector along the direction of oscillation, respectively. The contraction of the Green's dyadic $\overleftrightarrow{G}(\mathbf{r}_\mu, \mathbf{r}_\mu, \omega)$ is obtained from the real part of the electric field. The above expression must be multiplied by the quantum efficiency, which we assume here to be 1.

The β factor is calculated from the identity $\beta = \gamma_{CPP}/\gamma = (\gamma_{CPP}/\gamma_0)/P_F$. Here $\gamma$ and $\gamma_0$ are the total decay rates of the emitter inside the VG and in vacuum, respectively, whereas $\gamma_{CPP}$ is the decay rate into the guided modes. For the case in which there is no diamond host, the ratio $\gamma_{CPP}/\gamma_0$ can be obtained directly from the field profile of the VG eigenmode[33]. The final expression for the beta factor is thus: $\beta = \frac{\lambda_0^2}{4\pi c \mu_0 P_F} \frac{|\mathbf{u}_\mu \cdot \mathbf{e}(\mathbf{r}_\mu)|^2}{Re \int dA (\mathbf{e} \times \mathbf{h}^*)}$, where $\mathbf{e}$ and $\mathbf{h}$ are the electric and magnetic transverse fields of the eigenmode, and $\mu_0$ the vacuum permeability. The surface integral in the denominator spans over the transverse plane (xz), where the vector $\mathbf{u}_\mu$ is assumed to be contained.

When the diamond host is introduced, however, the system (VG + host) ceases to be translational invariant, and a different expression for the β-factor has to be used. In this case, we will determine it by the ratio $\beta = W_{CPP}/W$, where $W$ is the total power emitted by the NV inside the VG, and $W_{CPP}$ is the total power emitted into the modes of the VG. The former can be determined from our previously calculated Purcell factor, via $W = P_F W_0$. The power emitted by a dipole radiating in vacuum, $W_0$, is well known from the literature. To calculate the total power coupled to the CPP, we compute the overlap of the fields of the 3D problem, $\mathbf{E}$, and the fields of the VG eigenmode, $\mathbf{h}$.

For convergence reasons it is better to evaluate this overlap at a cross section outside the diamond host, located at a longitudinal distance $y_0$ away from the dipole. It is then necessary to include a compensation factor $\exp(2y_0 Im[\kappa])$ which takes into account the propagation losses of the CPP. The final expression for the β-factor is thus: $\beta = \frac{1}{W} e^{2y_0 Im[\kappa]} \langle \mathbf{E}(y_0) | \mathbf{h}(y_0) \rangle = \frac{1}{P_F W_0} e^{2y_0 Im[\kappa]} \frac{|\int dA (\mathbf{E} \times \mathbf{h}^*)|^2}{Re[\int dA (\mathbf{e} \times \mathbf{h}^*)]}$.

**Acknowledgements.** EBU, RM, MG and RQ acknowledge the European Community's Seventh Framework Programme (grant ERC- Plasmolight; no. 259196) and Fundació privada CELLEX. EBU acknowledges support of the FPI fellowship from the Spanish MICINN. R.M. acknowledges support of Marie Curie and NEST fellowships. CGB and FJGV acknowledge the European Research Council (ERC-2011-AdG, Proposal No. 290981). CGB, EM, and FJGV acknowledge the Spanish MINECO (Contract No. MAT2011-28581-C02-01). CGB acknowledges support of the FPU fellowship from the Spanish MECD. IPR, TH and SIB acknowledge financial support for this work from the Danish Council for Independent Research (the FTP project ANAP, Contract No. 09-072949) and from the European Research Council, Grant No. 341054 (PLAQNAP). YA acknowledges the support of RYC-2011-08471 fellowship from MICINN. The authors thank Luis Martin-Moreno and Cesar E. García for fruitful discussions, and Jana M. Say and Louise J. Brown for providing the ND solution, and Ioannis Tsioutsios for support with the AFM manipulation technique.

**Author contributions**
SIB and RQ conceived the experiment. CGB carried out the theoretical simulations. CGB, EM and FJGV analysed the theoretical results. MG built the experimental setup. TH and IPR fabricated and characterized the VGs. YA developed the electrostatic assisted assembly technique. EBU positioned the NDs and assembled the hybrid system. EBU, MG and RM performed the experimental measurements. EBU processed the experimental data. EBU and RM co-wrote the manuscript. All authors discussed the results and commented on the manuscript. SIB, FJGV and RQ supervised the project.

**Competing financial interests:** The authors declare no competing financial interests.